\documentclass[11pt]{article}
\usepackage{latexsym}
\newcommand{\pl}{{\it Phys.\ Lett. }}
\newcommand{\np}{{\it Nucl.\ Phys. }}
\newcommand{\prl}{{\it Phys.\ Rev.\ Lett. }}
\newcommand{\pr}{{\it Phys.\ Rev. }}

\newcommand{\xxx}[1]{ [#1]}
\newcommand{\mysection}[1]{\section{#1}
\setcounter{equation}{0}}

\def\thefootnote{\fnsymbol{footnote}}

\def\tg{{\tilde g}}

\def\tH{{\tilde H}}

\def\tB{{\tilde B}}

\def\[{\left [}
\def\]{\right ]}
\def\({\left (}
\def\){\right )}
\def\|{\left |}

\def\pp{\partial}

\def\Tr{{\rm Tr}}

\def\im{{\rm Im}}
\def\re{{\rm Re}}

\def\GeV{{\rm GeV}}

\newcommand{\beq}{\begin{equation}}
\newcommand{\eeq}{\end{equation}}
\newcommand{\bea}{\begin{eqnarray}}
\newcommand{\eea}{\end{eqnarray}}

\def\L{{\cal L}}

\def\pp{\partial}

\def\Del{\Delta}

\newcommand{\half}{{1 \over 2}}

\def\tq{{\tilde q}}

\newcommand{\mysec}[1]{Section \ref{#1}}

\begin{document}
\begin{titlepage}

\hfill   LBNL-57549

\hfill   UCB-PTH-08/11

\hfill   hep-th/yymmnnn


\begin{center}

\vspace{18pt}
{\bf Supersymmetry and Superstring Phenomenology}\footnote{This
work was supported in part by the
Director, Office of Science, Office of High Energy and Nuclear
Physics, Division of High Energy Physics of the U.S. Department of
Energy under Contract DE-AC03-76SF00098, in part by the National
Science Foundation under grant PHY-0457315.}\footnote{To be published
in ``Supersymmetry on the Eve of the LHC'', a special volume of
the European Physical Journal C (EPJC) dedicated to the
memory of Julius Wess.}

\vspace{18pt}

Mary K. Gaillard {\em and} Bruno Zumino
\vskip .01in
{\em Department of Physics, University of California 
and \\ Theoretical Physics Group, Bldg. 50A5104,
Lawrence Berkeley National Laboratory \\ Berkeley,
CA 94720 USA}

\vspace{18pt}

\end{center}

\begin{abstract} We briefly cover the early history of supersymmetry,
  describe the relation of SUSY quantum field theories to superstring
  theories and explain why they are considered a likely tool to
  describe the phenomenology of high energy particle theory beyond the
  Standard Model.

\end{abstract}

\end{titlepage}

\newpage
\null
\newpage
\renewcommand{\thepage}{\roman{page}}

\begin{center} SUPERSYMMETRY AND SUPERSTRING PHENOMENOLOGY

Mary K. Gaillard and Bruno Zumino\\
University of California at Berkeley\end{center}

\renewcommand{\thepage}{\arabic{page}}
\setcounter{page}{1}

\def\thefootnote{\arabic{footnote}} \setcounter{footnote}{0}
\mysection{Introduction} 
This paper is dedicated to the memory of Julius Wess, one of the
inventors of supersymmetry (SUSY), who passed away in August 2007.
Julius participated in the conference SUSY 07, that took place in
Karlsruhe, and gave an invited lecture there, just a few weeks before
his death.  In his lecture he detailed a brief history of the discovery
of SUSY quantum field theories in four space-time dimensions, and
expressed his hope that the superparticles predicted by SUSY will be
found in experiments at the Large Hadron Collider (LHC) at CERN that are
expected to begin this summer.

In this paper we will briefly cover the early history of SUSY, describe
the relation of SUSY quantum field theories to superstring theories
and explain why they are considered a likely tool to describe the
phenomenology of high energy particle theory beyond the Standard Model
(SM).  More extensive descriptions of SUSY and its applications to
physics can be found in the now standard text by Wess and Bagger~\cite{wb}
as well as in several recently published books~\cite{books}.

\mysection{The beginning of supersymmetry}\label{susy}
Four dimensional SUSY was discovered independently three times in the
early 1970's: first in Moscow by Golfand and Likhtman, then in
Kharkov, by Volkov and Akulov, and Volkov and Soroka, and finally by
Julius Wess and one of us (BZ), who collaborated at CERN in Geneva and
in Karlsruhe, and were totally unaware of the previous work.  It is
remarkable that Volkov and his collaborators did not know about the
work of Golfand and Likhtman, since all of them were publishing papers
in Russian in Soviet Journals.  For information on the life and work
of Golfand and Likhtman, we refer to the Yuri Golfand Memorial Seminar
Volume~\cite{golf}. This book contains an interesting graph on page 43
that shows the remarkably fast increase in the number of papers on
SUSY as a function of time after the first three {\it preprints} that
carried the names of Wess and Zumino.  For information on Volkov's
life and work, we refer to the Proceedings of the 1997 Volkov Memorial
Seminar in Kharkov~\cite{volk}.

Supersymmetry is a symmetry that relates the properties of
integral-spin bosons to those of half-integral spin fermions.  The
generators of the symmetry form what has come to be called a
superalgebra, which is a super extension of the Poincar\'e Lie algebra
of quantum field theory (Lorentz transformations and space-time
translations) by fermionic generators.  In a superalgebra both
commutators and anticommutators occur.  The early physics papers on
SUSY also gave rise to renewed interest by mathematicians in the
theory of superalgebras.  Eventually a complete classification of
simple and semisimple superalgebras was obtained, analogous to
Cartan's classification of Lie algebras, and even the prefix super was
adopted in mathematics.  Unfortunately the Poincar\'e superalgebra is
not semisimple, although it can be obtained by a suitable contraction;
the situation is similar to that for the Poincar\'e Lie algebra.  The
general classification of superalgebras does not appear to be very
useful in physics, because, unlike Lie algebras, simple or semisimple
superalgebras cannot be used as internal symmetries, or so it seems.

The early work on supersymmetric field theories considered only one
fermionic generator which is a Majorana spinor.  The corresponding
superalgebra is therefore called $N=1$ SUSY.  An important development
was the study of extended ($N>1$) SUSY and the construction of quantum
field theories admitting extended SUSY.  It turns out that $N=1$ SUSY
in four space-time dimensions is still the best choice for a SUSY
extension of the Standard Model of elementary particle physics,
because of the chirality properties of physical fermions. We describe
in \mysec{mssm} a popular version of such an extension: the Minimal
Supersymmetric Standard Model (MSSM).

\mysection{Superspace}\label{sugra}
The influence of SUSY on mathematics can be seen by the great interest
mathematicians have developed in the study of supermanifolds. From a
physicist's point of view this began with an important paper by A.
Salam and J. Strathdee~\cite{ss} who introduced the concept of ``superspace'',
a space with both commuting and anticommuting coordinates, and showed that
$N=1$ supersymmetry can be defined as a set of coordinate transfromations 
in supespace. Ferrara, Wess and Zumino then wrote some papers using the concept
of ``superfields'' (fields in superspace). Eventually the technique of 
superpropagators was developed and shown to be a useful tool for supersymmetric
perturbation theory.

With the discovery of supergravity (SUGRA), it became natural to study the
geometry of curved supermanifolds.   Wess and Zumino realized that superRiemannian
geometry had to be enlarged by the introduction of  a super-vielbien and a 
constrained, but nonvanishing, supertorsion.  They also formulated the geometry
in terms of exterior superdifferential forms, not unlike those introduced
independently by Berezin.  

Just as in the case of supersymmetric quantum field theories in flat
space, SUGRA has remarkable properties of convergence; however it does
not provide a theory that is finite or even renormalizable.  $N=1$
SUGRA admits higher $N$ extensions.  For various reasons, the highest
physically acceptable value is $N=8$,  Recently, remarkable properties
of $N=8$ SUGRA in four dimensions have been discovered~\cite{zvi}.
With clever combinations of diagrams, some of the perturbation theory
divergences cancel; {\it these} cancellations do not seem to be understandable
from the supersymmetry of the theory.  Are they due to some other symmetry?
Is $N=8$ SUGRA in 4 dimensions actually finite in perturbation theory?

\mysection{Successes and problems of the Standard Model}\label{sm}
The Standard Model can be briefly summarized as follows:\newline
Spin-${1\over2}$ fermions, called quarks $q$ and leptons $\ell$
fall into three families with identical interactions but increasing mass
(except for the neutrinos whose masses are all very small):
\bea 
SU(3)&&\;\;\qquad\Longleftrightarrow\;\;\qquad\qquad\qquad\qquad
\Longleftrightarrow\quad\qquad\qquad\qquad\qquad\Longleftrightarrow 
\nonumber\\
&&\pmatrix{u_1&u_2&u_3 &\nu_e\cr d_1&d_2&d_3 &e\cr}\qquad
\pmatrix{c_1&c_2&c_3 &\nu_\mu\cr s_1&s_2&s_3&\mu\cr}\qquad
\pmatrix{t_1&t_2&t_3 &\nu_\tau\cr b_1&b_2&b_3 &\tau\cr}\qquad\Updownarrow\quad
SU(2)\nonumber\eea
The labels $u,d,c,s,t,b$ for quarks $q$ and $e,\mu,\tau$ for negatively
charged leptons $\ell^-$ and their associated neutrinos $\nu_\ell$ distinguish
among the fermion ``flavors'', and the indices $i=1,2,3$ for quarks of fixed
flavor distinguish among three ``colors''.
Interactions among fermions are mediated by 
spin-1 gauge bosons $A_\mu$:
\bea SU(3)&&\quad\Longleftrightarrow\nonumber\\ &&
g_1,\ldots,g_8\qquad\qquad \pmatrix{W_1\cr W_2\cr
W_3}\quad\Updownarrow\quad SU(2) \qquad\qquad B \nonumber\eea
where the eight gluons $g$ couple quarks of different color but the
same flavor, and the three $W$'s couple fermions of different flavor .
The laws of physics are invariant under local $SU(3)\otimes
SU(2)\otimes U(1)$ transformations, with $SU(3)$ acting horizontally
on color and SU(2) acting vertically on flavor as indicated by the
arrows.  The vacuum configuration is not invariant under the full
$SU(3)\otimes SU(2)\otimes U(1)$ group of symmetries; ``spontaneous''
breaking of the electroweak symmetry down to gauge invariance of
electromagnetism (em)
$$ SU(2)\otimes U(1)\to U(1)_{\rm em} $$
is attributed to the vacuum value of the spin-0 ``Higgs'' field $H$
$$v = \langle{H}\rangle\approx 250\GeV$$
which gives masses to the electrically charged $W$'s and one linear
combination of the neutral $W$ and the $U(1)$ gauge boson $B$:
 $$Z =
\cos\theta_w W_3 - \sin\theta_w B,\qquad \gamma = \cos\theta_w B +
\sin\theta_w W_3,$$ 
$$m_\gamma = 0,\qquad m_{W_{1,2}} = {g\over{2}}v =
m_Z\cos\theta_w,$$ 
where $g$ is the $SU(2)$ coupling constant.  The fermions also acquire
masses through their coupling to the Higgs field which also induces a
small mixing among fermions of the same electric charge but different
flavors, with mixing amplitudes of order:
$$\nu_e\leftrightarrow\nu_\mu\leftrightarrow \nu_\tau\ll
d\leftrightarrow s \leftrightarrow b\ll 1.$$
The absence of large flavor changing effects is an important
property of the Standard Model that has to be reproduced by the
underlying theory.

The SM has been tested to a high degree of accuracy, but it is not
without its difficulties. Foremost among these is the large hierarchy
between the electroweak symmetry breaking scale and the reduced
Planck scale $m_P$, known as the ``gauge hierarchy'':
$$ m_Z\approx 90\GeV \ll m_{\rm P} = \sqrt{8\pi\over
G_{\rm Newton}}\approx2\times10^{18}\GeV,$$
which is hard to understand within the context of quantum field
theory.  In addition there is no understanding of the pattern of quark
masses and flavor mixing, nor why the $\theta$-parameter, that governs
CP violation in the QCD vacuum, is so small: $\theta< 10^{-10}$, where
QCD, or quantum chromodynamics, is the $SU(3)$ gauge theory of
the Standard Model. The only potential SM candidate for dark matter, a
neutrino, has been ruled out by the study of galaxy formation.  On the
other hand, many extensions of the SM that attempt to address the
``$\theta$-problem'' require the existence of a very light
pseudoscalar, or axion, that remains a viable dark matter candidate.
The issues of the cosmological constant and dark energy are totally
outside the domain of the SM.

\mysection{The particle content of the MSSM}\label{mssm} 
One way to understand the gauge hierarchy alluded to above is through
supersymmetry: cancellations among boson and fermion loops remove
ultraviolet divergences and stabilize mass hierarchies against large
quantum corrections.  This requires a doubling of the number of
particles in the SM.  Spin-${1\over2}$ quarks and leptons form chiral
supermultiplets with spin-0 squarks $\tilde q$ and sleptons
$\tilde\ell$:
\bea
SU(3)&&\;\;\qquad\Longleftrightarrow\;\;\;\qquad\qquad\qquad\qquad
\Longleftrightarrow\;\;\qquad\qquad\qquad\qquad\Longleftrightarrow \nonumber\\
&&\pmatrix{\tilde u_1&\tilde u_2&\tilde u_3 &\tilde \nu_e\cr\tilde
d_1&\tilde d_2&\tilde d_3 &\tilde e\cr}\qquad \pmatrix{\tilde
c_1&\tilde c_2&\tilde c_3 &\tilde \nu_\mu\cr\tilde s_1&\tilde
s_2&\tilde s_3&\tilde \mu\cr}\qquad \pmatrix{\tilde t_1&\tilde
t_2&\tilde t_3&\tilde \nu_\tau\cr\tilde b_1&\tilde b_2&\tilde
b_3 &\tilde \tau\cr}\qquad\Updownarrow\quad SU(2)\nonumber\eea
and spin-1 gauge bosons $A_\mu$ form vector supermultiplets with
spin-${1\over2}$ gauginos $\lambda$:
\bea SU(3)&&\quad\Longleftrightarrow\nonumber\\ && \tilde
g_1,\ldots,\tilde g_8\qquad\qquad \pmatrix{\tilde W_1\cr\tilde
W_2\cr\tilde W_3} \quad\Updownarrow\quad SU(2) \qquad\qquad \tilde B
\nonumber\eea
called gluinos, Winos and Bino, respectively.
There is also the Higgsino fermionic superpartner of the SM Higgs boson, but
the supersymmetric extension of the the SM requires {\it two} Higgs
chiral supermultiplets in order to avoid gauge anomalies, generate
mass terms for all charged chiral fermions as well as for all Winos.
Thus the spectrum includes
\bea \pmatrix{H_u^+&\tH_u^+&H_d^0&\tH_d^0\cr
H_u^0&\tH_u^0&H_d^-&\tH_d^-\cr} \quad\Updownarrow\quad SU(2)\nonumber\eea
and their charge conjugates. While there is no experimental evidence
for SUSY partners--indeed there are ever increasing lower bounds on
their masses--there is a tantalizing piece of indirect evidence.  In
the SM, the three coupling constants of the $SU(3)\otimes SU(2)\otimes
U(1)$ gauge theory almost converge at a single point at an energy of
about $10^{15}\GeV$, but present high precision data excludes exact
unification by about 9 standard deviations.  In the MSSM the running
of the coupling constants is modified, most importantly by the
presence of {\it two} Higgs {\it super}multiplets, which together are
equivalent to six ordinary Higgs fields, and their convergence is much
closer, now at an energy of about $10^{16}\GeV$.  The higher scale of
unification is welcome in view of strong limits on the proton
lifetime, since theories that unify the SM gauge groups generally
predict proton decay with a Fermi coupling inversely proportional to
the square of the unification scale.  In addition it is just two
orders of magnitude below the reduced Planck scale, perhaps a hint at
unification with gravity?

However the MSSM is not without problems of its own.  As explained in
\mysec{bmssm} supersymmetry cannot be broken spontaneously, requiring
the introduction of a large number of arbitrary complex parameters and
no rationale for suppressing flavor and CP violating terms to the
small values needed to be consistent with observation.  In addition
one has to impose a discrete symmetry, called R-symmetry, to avoid
fast proton decay and suppress lepton flavor violating processes such
as $\mu\to3e$.  Thus one assigns $R = +1$ to the particles of the SM
including $H_u,H_d$, and $R = -1$ to their superpartners. As a
consequence superpartners must be produced in pairs, and the lightest
one is stable, providing a possible dark matter candidate, as
discussed in \mysec{dark}.

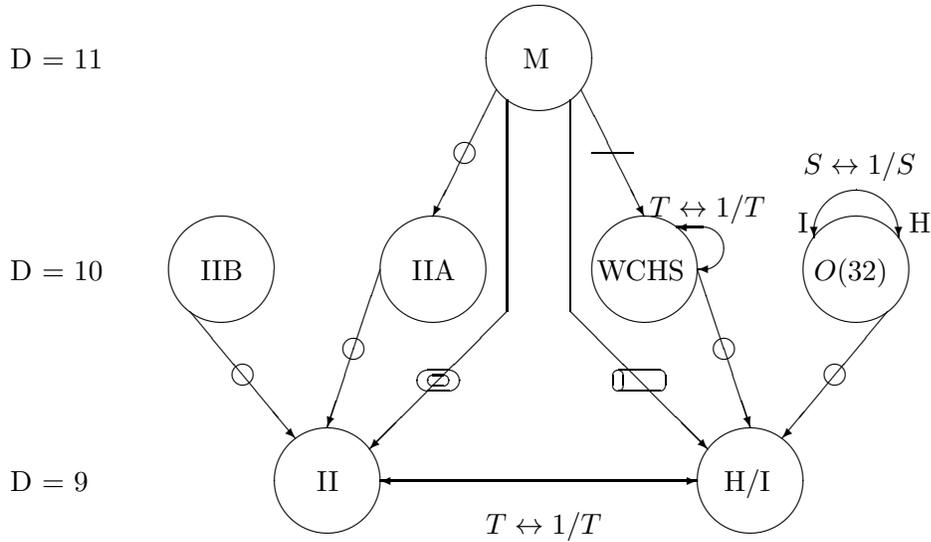
\begin{figure}
\begin{picture}(350,220)(0,20)
\put(0,36){D = 9}\put(0,116){D = 10}\put(0,196){D = 11}
\multiput(120,40)(160,0){2}{\circle{40}}
\multiput(80,120)(80,0){4}{\circle{40}} \put(200,200){\circle{40}}
\put(68,104){\line(5,-6){40}}\put(260,120){\vector(1,-3){20}}
\put(107,57){\vector(1,-1){1}}\put(293,57){\vector(-1,-1){1}}
\put(332,104){\line(-5,-6){40}}\put(140,120){\vector(-1,-3){20}}
\put(184,188){\vector(-1,-2){24}}\put(216,188){\vector(1,-2){24}}
\put(188,184){\line(0,-1){80}}\put(212,184){\line(0,-1){80}}
\put(188,104){\vector(-1,-1){52}}\put(212,104){\vector(1,-1){52}}
\put(200,40){\vector(-1,0){60}}\put(200,40){\vector(1,0){60}}
\put(262,120){\vector(-1,0){2}}\put(262,136){\vector(-1,0){10}}
\put(304,134){\vector(0,-1){2}}\put(336,134){\vector(0,-1){2}}
\put(262,128){\oval(16,16)[r]}\put(320,134){\oval(32,32)[t]}
\put(222,116){WCHS}\put(304,116){$O(32)$}
\put(298,134){I}\put(340,134){H}\put(194,196){M}
\put(116,36){II}\put(270,36){H/I}\put(72,116){IIB}\put(152,116){IIA}
\put(88,80){\circle{8}}\put(130,90){\circle{8}}\put(220,164){\line(1,0){16}}
\put(312,80){\circle{8}}\put(270,90){\circle{8}}\put(172,164){\circle{8}}
\put(162,78){\oval(8,4)}\put(162,78){\oval(16,8)}
\put(230,78){\oval(4,8)}\put(246,78){\oval(4,8)[r]}
\put(230,82){\line(1,0){16}}\put(230,74){\line(1,0){16}}
\put(180,20){$T\leftrightarrow 1/T$}\put(242,140){$T\leftrightarrow
1/T$} \put(300,156){$S\leftrightarrow 1/S$}
\end{picture}
\caption{M-theory according to John Schwarz.
\label{fig:john}}
\end{figure}

\mysection{Supersymmetry in string theory}\label{string}
When supersymmetry is combined with general relativity, one gets
gauged supersymmetry, or supergravity, which is the infinite
string tension limit of string theory. 
At present most particle theorists view superstring theory in 10
dimensions as the most promising candidate that reconciles general
relativity with quantum mechanics. There are five of these theories,
and they are related to one another by two types of dualities:
S-duality that relates strong to weak coupling: $\alpha\to 1/\alpha,$
and T-duality that relates large to small radius of compactification:
$R \to 1/R$, where $\alpha= g_s^2/4\pi$ is the fine structure constant
of the gauge group(s) at the string scale, and $R$ is a radius of
compactification from dimension D to dimension ${\rm D} -1$.
Figure~\ref{fig:john} shows~\cite{john} how these dualities relate the
various 10-D superstring theories to one another, and to M-theory,
which lives in 11 dimensions and involves membranes.  In
Figure~\ref{fig:john} the small circles, line, torus and cylinder
represent the relevant compact manifolds in reducing D by one or two.
The two $O(32)$ theories are S-dual to one another, while the
$E_8\otimes E_8$ weakly coupled heterotic string theory (WCHS) is
perturbatively invariant~\cite{mod} under T-duality when compactified
to four dimensions; in this case $R$ is the radius of the compact
six dimensional manifold. 

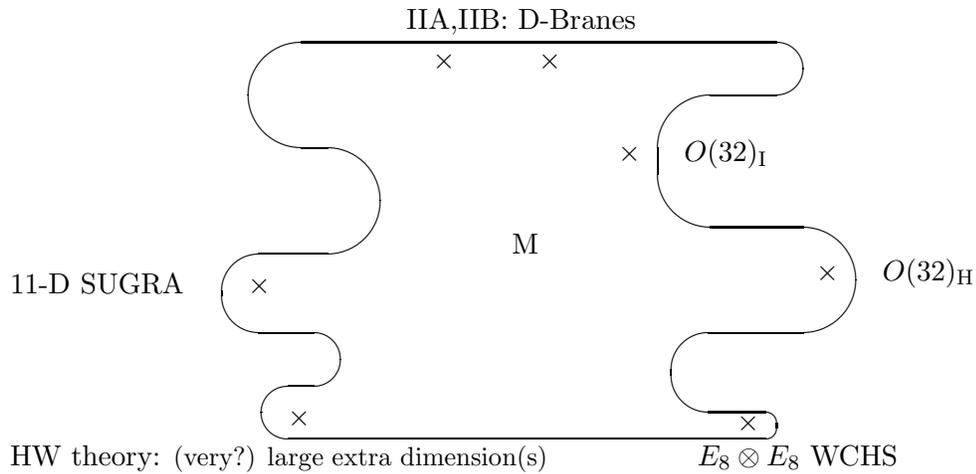
\begin{figure}
\begin{picture}(350,215)(0,40)
\put(150,205){IIA,IIB: D-Branes}\put(120,200){\line(1,0){160}}
\multiput(160,190)(40,0){2}{$\times$}
\put(120,180){\oval(60,40)[l]}\put(280,190){\oval(40,20)[r]}
\put(120,140){\oval(40,40)[r]}\put(280,155){\oval(70,50)[l]}
\put(120,120){\line(-1,0){10}}\put(110,105){\oval(60,30)[l]}
\put(110,80){\oval(30,20)[r]}
\put(110,60){\oval(30,20)[l]}\put(280,110){\oval(80,40)[r]}
\put(280,90){\line(-1,0){10}}\put(270,75){\oval(40,30)[l]}
\put(270,55){\oval(40,10)[r]}\put(110,50){\line(1,0){160}}
\put(230,155){$\times$}\put(255,155){$O(32)_{\rm I}$}
\put(305,110){$\times$}\put(330,110){$O(32)_{\rm H}$}
\put(90,105){$\times$}\put(0,105){11-D SUGRA}\put(190,120){M}
\put(275,53){$\times$}\put(260,40){$E_8\otimes E_8$ WCHS}
\put(105,55){$\times$}\put(0,40){HW theory: {\small (very?) 
large extra dimension(s)}}\end{picture}
\caption{M-theory according to Mike Green.
\label{fig:mike}}\end{figure}

We will be specifically focusing on this
theory, and T-duality will play an important role.
Another image of M-theory, the ``puddle diagram'' of
Figure~\ref{fig:mike}, indicates~\cite{mike} that all the known
superstring theories, as well as \mbox{D $=11$} supergravity, are
particular limits of M-theory.  Each point in the puddle has a very
large number of possible vacua, and currently there is a lot of
activity in trying to count the number of type IIB vacua; the number
is very large. This endeavor is related to an attempt to address the
problem of the cosmological constant as mentioned in \mysec{dark} and
discussed in~\cite{bruno}. The Ho\v rava-Witten (HW) scenario~\cite{hw} and its
inspirations have also received considerable attention.  If one
compactifies one dimension of the 11-dimensional M-theory, one gets
the HW scenario with two 10-D branes, each having an $E_8$ gauge
group.  As the radius of this 11th dimension is shrunk to zero, the
weakly coupled heterotic string scenario is recovered; it is this
string theory that most naturally incorporates the standard model.
In the limit of infinite string tension, the WCHS reduces to 10-D
supergravity coupled to an $E_8\otimes E_8$ Yang-Mills gauge
supermultiplet.  If 6 dimensions are compactified on, say, three
2-tori, that have a flat geometry, the resulting 4-D theory will have
N=4 supersymmetry because the 8-component spinor that generates
supersymmetry transformations in 10 dimensions gives four 2-component
supersymmetry generators in four dimensions. As explained in
\mysec{susy} only N=1 supersymmetry can provide a viable framework for
observable particle physics.  Therefore we need a curved 6-D manifold
that has a nontrivial holonomy group (group of transformations under
parallel transport).  Calabi-Yau compactification has
an $SU(3)\in SO(6)$ holonomy group that leaves only one 2-component
spinor single-valued under parallel transport on the compact manifold,
giving N=1 supersymmetry in four dimensions. The gauge group may be
decomposed as
$$E'_8\otimes E_8\ni E'_8\otimes E_6\otimes SU(3),$$
and the equations of motion require that the space-time background
curvature be balanced against a background gauge field strength in
such a way that an $SU(3)$ subgroup of one $E_8$ is identified with
the $SU(3)$ holonomy group of the Calabi-Yau manifold.  Then the
surviving 4-D gauge group is $E'_8\otimes E_6$.  Orbifold
compactifications that mimic the Calabi-Yau case have been studied
more extensively because the compact manifold is flat except at
singular points with infinite curvature, and therefore it is easier to
extract the low energy theory.  In this case the residual gauge group
in four dimensions is
$$E'_8\otimes E_6\otimes [G'\in SU(3)].$$
In either case, the outcome is promising because $E_6$ has long been
recognized as the largest group that is a viable candidate for the
unification of the strong and electroweak gauge groups of the Standard
Model.  The massless spectrum consists of single-valued fields that
are invariant under the diagonal of the two broken $SU(3)$'s (or, in
orbifold compactification, an appropriate subgroup thereof). The
surviving degrees of freedom of the 10-D gauge supermultiplet
$$(A_M,\tg)_{E'_8\otimes E_8}, \qquad M = 0,\ldots,9,$$
are the 4-D gauge supermultiplet
$$ (A_\mu,\tg)_{E'_8\otimes E_6}, \quad  \quad \mu = 0,\ldots,3,$$
which is invariant under both $SU(3)$'s, and matter chiral multiplets
$$(A_m,\tg)_{E_8/[E_6\otimes SU(3)]} = (27 +
\overline{27})_{E_6},\qquad m = 4,\ldots,9,$$
which transform as $(3,\bar 3) + (\bar 3,3)$. These states decompose
under the smaller candidate gauge unification groups $SO(10)$ and $SU(5)$
as
$$ 27_{E_6} = (16 + 10 + 1)_{SO(10)} = (\bar{5} + 10 + 1 + 5 + \bar{5} 
+ 1)_{SU(5)}.$$
The $\bar 5 + 10$ of $SU(5)$ contains the quarks and leptons of the
standard model. These form the 16 of $SO(10)$ together with a
Standard Model singlet that may be responsible for the recently
observed small neutrino masses and mixing. The $5 + \bar 5$ contained
in the 10 of $SO(10)$ includes, among other things, the two Higgs
doublets needed in the supersymmetric extension of the Standard Model,
as discussed in \mysec{mssm}.
What does not appear in the massless spectrum is a chiral multiplet
transforming according to a large representation of the gauge group, such as
the adjoint representation, that could include a Higgs particle whose
vacuum value would break $E_6$ to the Standard Model. Instead this is achieved 
by what is known as the Hosotani mechanism or Wilson lines.
If the compact manifold is not simply connected, gauge flux can be 
trapped around a noncontractible loop:
$$\left<\int dl^m A_m\right> \ne 0,$$
in a manner reminiscent of the Arahonov-Bohm effect.   The nonvanishing
gauge flux has the same effect as an adjoint Higgs field, further
breaking the symmetry to leave a 4-D gauge group:  
$$(G_{\rm hid}\in E'_8)\otimes[(SM\otimes G'')\in E_6]\otimes [G'\in SU(3)]$$
where the Standard Model gauge group is SM $=SU(3)\otimes SU(2)\otimes
U(1)$. As mentioned previously, the class of vacua described here is a
tiny subset of the full set of possible vacua, even within the
framework of the WCHS; a recent review of the many facets of  string theory
and M-theory can be found in \cite{b2s}. The attractiveness of Calabi-Yau-like
compactification of the WCHS is that the gauge group and
states of the SM emerge naturally.  In addition the spectrum includes
an axion that is a candidate for the ``QCD axion'' mentioned at the end of
\mysec{sm} and/or for dark matter.  The presence of a hidden sector is
also welcome, as it may provide a mechanism for spontaneous
supersymmetry breaking.

\mysection{Dark matter and dark energy in supersymmetry}\label{dark}
Many independent lines of cosmological evidence have led to the
conclusion that the vast majority of matter in the universe is
``dark'' in the sense that it has evaded observation based on direct
interaction with electromagnetic radiation.  Nonbaryonic dark matter
out-masses ordinary matter by a factor of about 8.  The dominant class
of dark matter candidates are ``Weakly Interacting Massive Particles''
(WIMPs).  There have been a number of suggestions for dark matter
particles, but it seems that the best candidate is the lightest
``neutralino'' that is provided by TeV-scale SUSY.

A particle dark matter candidate must satisfy the following criteria:
\begin{itemize}
\item It must be ``stable'' in the sense that its lifetime is longer than 
the age of the universe.
\item There must be an effective production mechanism to create the right
amount in the early universe.
\item It must be nonrelativistic during structure formation; in other words
it must be ``cold'' dark matter.
\item It must be weakly interacting to have escaped detection, which
implies that it must be electrically neutral and colorless.
\end{itemize}
These constraints are satisfied by the neutral Higgsinos
($\tH_u,\tH_d$).  the neutral Wino ($\tilde W^0$) and the Bino
($\tB$), four Majorana fermions with the same quantum numbers, that
can mix to give four mass eigenstates:
$\chi^0_1,\chi^0_2,\chi^0_3,\chi^0_4,$ that are called neutralinos.
The lightest of these is stable and a good candidate for dark matter.  Other
possibilities include the lightest sneutrino, which is apparently
excluded by accelerator experiments, and the superpartner of the
graviton, the gravitino, which would be very hard to detect. A very
comprehensive review of cold dark matter can be found in~\cite{silk}.

In addition, it appears to be generally accepted by astrophysicists
and cosmologists that the cosmological constant $\Lambda$, long
believed to vanish, is actually positive but very small.  As a
consequence, the expansion of the universe is accelerating.  If we
interpret $\Lambda$ as the energy density of the vacuum, dimensional arguments,
as well as quantum field theory (QFT) calculations would give it a value of
$$\Lambda_P = {\rm {Planck\; mass\over(Planck\;length)^3}}
\approx 10^{94}{\rm {grams/cm^3}},$$
while the actual value is $\Lambda\sim 10^{-120}\Lambda_P$.

In SUSY QFT, without supergravity, the vacuum energy vanishes to all
order in perturbation theory.  In a generic QFT it diverges
quartically, while if SUSY is broken only softly it diverges
logarithmically.  Still, for any reasonable cut-off, $\Lambda$
comes out much larger than the measured value.  The MSSM and other
SUSY extension of the SM have nothing to say about this.  In
supergravity, the cosmological constant is no longer related to the
scale of supersymmetry breaking and can take any value, but there is
no simple understanding of its measured value. As mentioned in
\mysec{string}, there is considerable activity in counting vacua with
the notion that there might be a probabilistic determination of the
cosmological constant; this raises a number of philosophical questions
that we do not wish to enter into here.

\mysection{Breaking supersymmetry: beyond the MSSM}\label{bmssm}
The absence of observed superpartners implies that supersymmetry is
broken in the vacuum. Furthermore an analysis of the observed particle
spectrum shows that supersymmetry cannot be spontaneously broken in
the observable sector. This can simply be seen as follows~\cite{cc}.
The squark squared mass matrices $M^2_{|Q|}$, are $6\times6$ matrices
in flavor space; $Q$ is the electric charge of the squark sextuplets
$(\tilde u_L,\tilde c_L, \tilde t_L,\tilde u_R,\tilde c_R, \tilde
t_R),$ with $Q={2\over3}$, and $(\tilde d_L,\tilde s_L, \tilde
d_L,\tilde d_R,\tilde s_R, \tilde b_R)$, with $Q={-{1\over3}}$.  The
subscripts $L,R,$ denote, respectively, squarks in chiral
supermultiplets $\Phi(\tq_L,q_L)$ together with left-spinning quarks,
and squarks in antichiral supermultiplets
$\Phi^\dag(\tq_R,q_R)$ together with right-spinning quarks. If
supersymmetry is broken spontaneously and the vacuum conserves
electric charge and color, these matrices satisfy
$$ M^2_{2\over3} = \pmatrix{m^2_{2\over3} + \half g D^3_W +
{1\over6}g' D_Y & \Del\cr \Del^\dag& m^2_{2\over3} -
{2\over3}g' D_Y\cr}, \qquad
 M^2_{1\over3} = \pmatrix{m^2_{1\over3} - \half g D^3_W +
{1\over6}g' D_Y & \Del'\cr \Del'^\dag& m^2_{1\over3} +
{1\over3}g' D_Y\cr},$$
where $m^2_{|Q|}$ are the $3\times3$ squared mass matrices of the
corresponding quarks, the ``D-term" contributions are proportional
the appropriate electroweak quantum numbers of the squarks, and the
off-diagonal elements $\Del,\Del'$ do not need to be specified for
our purposes.  If we define a $12\times12$ squark squared  mass matrix
$M^2$, and a $6\times6$ quark squared mass matrix $m^2$, by
$$  M^2 = \pmatrix{M^2_{2\over3}&0\cr0& M^2_{1\over3}\cr},\qquad
m^2 = \pmatrix{m^2_{2\over3}&0\cr0& m^2_{1\over3}\cr},$$
We see immediately that $\Tr M^2 = 2\Tr m^2\approx 2m_t^2\approx
(240\GeV)^2 = \sum_{i=1}^{12}M^2_{\tilde q}$, while data from the
Tevatron at Fermilab puts a lower limit on the lightest squark mass at
about $250\GeV$.  However one might imagine that there is a fourth
family of quarks and leptons and their superpartners (with a heavy
neutrino, since a fourth light neutrino is ruled out by data from the
LEP collider at CERN and by the abundance of light elements).  This will
not help due to the following argument\cite{cc}. Since the D-terms on
the diagonal in $M^2$ sum to zero, either they are all zero, or at
least one of them is $\le0$.  Assume this is the case for the first one
$$\half g D_3 + {1\over6}g' D_Y = - a^2\le0.$$
Denote by $\beta$ the 3-component vector that is the eigenstate of
$m^2_{2\over3}$ with the smallest eigenvalue (i.e. the physical up
quark):  $m^2_{2\over3}\beta = m^2_u$. Then
$$ \bar\beta^\dag M^2_{2\over3}\bar\beta = m^2_u - a^2 \le m^2_u,
\qquad \bar\beta = \pmatrix{\beta\cr0\cr}.$$
But we know from matrix theory that $\bar\beta^\dag
M^2_{2\over3}\bar\beta$ is greater than or equal to the smallest
eigenvalue $M^2_0$ of $M^2_{2\over3}$: $M^2_0\le m^2_u$.  If one
assumes instead the second D-term $ - D_Y\le0$, one reaches the same
conclusion, and taking the third or fourth D-term to be $\le0$, one
finds $M'^2_0\le m^2_d$, where $M'^2_0$ is the smallest eigenvalue of
$M^2_{1\over3}$.  If SUSY is spontaneously broken in the MSSM there
must be a squark lighter then a few MeV, which is clearly ruled out
by electron-positron annihilation data.

Therefore the only way to break supersymmetry in a
supersymmetric extension of the Standard Model without reintroducing a gauge
hierarchy problem is by introducing ``soft'' (operators of dimension
three or less in the Lagrangian) supersymmetry breaking.  This leads
to a plethora of arbitrary parameters, and therefore to the idea that
supersymmetry must be spontaneously broken in a ``hidden sector'' of
the full theory.  For example there could be a sector that interacts with
ours only through gravitational strength couplings.  As mentioned in
\mysec{string}, this scenario arises naturally in the context if the
weakly coupled heterotic string.

For a gauge group $G_a$ the $\beta$-function, which governs the
energy-dependence of the coupling ``constant'' $g_a(\mu)$, is defined by
\bea \mu{\pp g_a(\mu)\over\pp\mu} &=& \beta(\mu) = 
-{3\over2}b_ag_a^3(\mu) + O(g^5_a).\nonumber\eea 
Suppose that the hidden sector gauge group $G_{\rm hid}$ contains a
subgroup $G_c$ with $\beta$-function coefficient $b_c$ that is larger
than the coefficient $b_{QCD}$ of the SM color gauge gauge theory,
QCD.  In this case hidden sector confinement and gaugino condensation
\beq\left\langle\lambda\lambda\right \rangle_{\rm hid} \ne 0\label{cond}\eeq
will occur at a scale $\Lambda_c$ exponentially larger than the scale
$\Lambda_{QCD}$ at which color confinement and quark condensation take
place in QCD.  To see how this can provide a source of supersymmetry
breaking in the superstring context, we first note that four
dimensional supergravity is specified by three functions of chiral
superfields: the superpotential $W(Z)$ which is a holomorphic function
that determines Yukawa couplings of chiral fermions to scalars, the
K\"ahler potential $K(Z,\bar Z)$ which is a real function that governs
the kinetic energy terms for chiral fields and the holomorphic
function $f(Z)$ whose vacuum value gives the gauge coupling and the
angle $\theta$ that determines the vacuum configuration of the
Yang-Mills fields: $\langle f(Z)\rangle = g^{-2} - i\theta/8\pi^2.$ In
addition we need to introduce important chiral supermultiplets, known
as moduli supermultiplets, that are remnants of 10-D supergravity.
The 10-D supergravity supermultiplet consists of the 10-D metric
$g_{MN}$, an antisymmetric tensor $B_{MN}$, a scalar $\phi$ known as
the dilaton, and their fermionic superpartners, the 10-D gravitino
$\psi_M$, and a spin-${1\over2}$ fermion $\chi$.  These are all
invariant under the gauge $SU(3)\in E_8$, so those that remain in the
massless spectrum must also be invariant under the holonomy $SU(3)\in
SO(6)$, leaving the graviton $g_{\mu\nu}$ and the gravitino $\psi_\mu$
of 4-D supergravity, and an invariant subset of the two real scalars
$g_{mn}$ and $B_{mn}$ and the spin-${1\over2}$ fermion $\psi_m$, the
4-D antisymmetric tensor $B_{\mu\nu}$, the dilaton $\phi$ and the
single valued component of $\chi$.  These combine to form chiral
supermultiplets, whose vacuum values determine the size and shape of
the compact manifold, the gauge coupling constant $g_s$ at the string
scale $m_s = g_s m_{\rm P}$, and the $\theta$ angle of the 4-D theory.
For example, in compactifications on an orbifold $M_6$ that factorizes
into three 2-tori $T^2$: $M_6 = (T^2/G)^3$, where $G$ is a discrete
group of symmetries on the torus, there are three ``K\"ahler moduli''
superfields $T^I$ with complex scalar components
$$t^I = \phi^{3\over 4}{\det}^{1\over2}g^I_{m n} + i{b^I_{12}\over\sqrt{2}},
\qquad \langle\re t^I\rangle = R^I,\qquad I = 1,2,3$$
where $g^I_{m n}$, $m,n=1,2$ is the 2-D metric on the $I$'th 2-torus and
$R^I$ is its radius. For some orbifolds there are also ``complex
structure'' moduli superfields $U^I$ with the scalar vacuum value
$\langle\re u^I\rangle$ determining the ratio of radii on the 2-torus.
More generally, there can be a $3\times 3$ matrix-valued modulus
superfield $T^{I J}$ in the low energy theory; however most phenomenological
studies assume that only the three diagonal moduli $T^I$ are part of
the massless spectrum.
All compactifications of the WCHS have a `` dilaton'' chiral
supermultiplet $S$ with $f(Z) = S$; 
thus the vacuum value of its scalar component $s$
determines the gauge coupling constant and the $\theta$ angle:
$$\left\langle s\right\rangle 
= g_s^{-2} - i\theta/8\pi^2.$$
The real part of its scalar component is given by
$$ \re s = \phi^{-{3\over 4}}{\det}g_{m n},$$
and the imaginary part, known as the ``universal axion'' is dual to
the 2-form:
$$ \pp_\mu\im s = \phi^{-{3\over2}}(\det g_{m n})
\epsilon_{\mu\nu\rho\sigma}\pp^\nu B^{\rho\sigma}.$$
As a consequence of the dilaton coupling to the Yang-Mills sector, gaugino
condensation (\ref{cond}) generates an 
effective superpotential for $S$
\beq W(S) \propto e^{-S/b_c},\label{ws}\eeq
which in turn induces a gravitino mass:
$$m_{3\over2}\propto\left\langle W(s)\right\rangle
\propto e^{-s/b_c} = e^{-1/b_cg^2_c} = \Lambda^3_c$$
thereby breaking local supersymmetry.
The form of the superpotential (\ref{ws}) led to what is called the
``runaway dilaton'' problem. At the classical level, the scalar
potential is proportional to the square of the superpotential:
$${V(s)\propto e^{-2\re s/b_c}},$$
so the vacuum corresponds to vanishing coupling and no condensation:
\mbox{$V(s)\to0$ for $\left<(\re s)^{-1}\right> = g^2\to 0$}.
However, the effective potential for $s$ is constructed by anomaly
matching~\cite{vy}:
$$\delta\L_{eff}(s) \longleftrightarrow\delta\L_{\rm hid}$$
under the classical symmetries of the effective QFT
that are anomalous at the quantum level. One of these is T-duality
itself, which is not anomalous in the underlying string theory.  The
symmetry is restored in the effective QFT by adding a four dimensional
counterpart~\cite{gs4} of the Green-Schwarz counter-term~\cite{gs} in
10-D supergravity.  This modifies the classical dilaton K\"ahler
potential: $K(s + \bar s) = -\ln(2\re s) \to K(\ell) = \ln\ell$, where
$\ell^{-1} = \[2\re s + b_{G S}\sum_I\ln(2\re t^I)\]$, and introduces
a second runaway direction to strong coupling: $ V(\ell) \to - \infty$
for $\langle\ell\rangle = g_s^2/2\to\infty$, where the weak coupling
approximation is no longer valid. Including~\cite{bgw} nonperturbative
string effects~\cite{shenk} \mbox{$\sim A e^{-c/\sqrt{\ell}}$} and/or
other corrections~\cite{bd} to the dilaton K\"ahler potential allows
for dilaton stabilization at weak coupling and very small vacuum
energy by adjusting, for example, the parameters $A$ and $c$ in a
parametrization of these effects in the region around the vacuum.
This mechanism for stabilizing the dilaton is known as ``K\"ahler''
stabilization.  The attractive features of K\"ahler stabilized, T
self-dual heterotic string models can be summarized as
follows~\cite{rev}
\begin{itemize} 
\item In contrast with models that stabilize the dilaton using more
than one gaugino condensate (and adjusting their relative
$\beta$-functions), there is no difficulty in generating a positive
semi-definite potential.
\item The K\"ahler moduli are stabilized at self-dual
  points~\cite{bgw2}: $T_{\rm s d}\to T_{\rm s d}$, with supersymmetry
  conserving vacuum values $\langle T_{\rm s d}\rangle$. As a
  consequence, no large flavor mixing is induced by supersymmetry
  breaking, which arises only from the condensate and dilaton vacuum
  values. In contrast to the K\"ahler moduli, the dilaton has
  flavor-independent couplings to observable matter.
\item The condition of (nearly) vanishing vacuum energy leads to mass
  hierarchies~\cite{bgw3}:
  $$m_\ell\gg m_t\sim(10-20) m_{3\over2}\sim m_0\gg m_{1\over2}$$
  The enhancement of the moduli masses $m_\ell,m_t$ relative to the
  gravitino mass $m_{3\over2 }$ avoids potential problems~\cite{bbn}
  with Big Bang nucleosynthesis, and the suppression of gaugino
  masses $m_{1\over2}$ relative to scalar masses $m_0$ results in
  important quantum corrections to the former.  As a result these
  models naturally accommodate a dark matter candidate~\cite{bn}.
\item One hidden sector condensate is sufficient to break
  supersymmetry and stabilize the dilaton.  As a result there is a
  residual R-symmetry (a continuous version of the R-parity introduced in
  \mysec{mssm}) that guarantees a massless axion at the QFT level.
\item In supergravity, this R-symmetry is protected by T-duality,
  allowing for a possible solution to the CP problem of QCD~\cite{bgk}.
\item T-duality provides a possible mechanism for generating R-parity
in the MSSM~\cite{mkg}.
\end{itemize}
The phenomenology of K\"ahler stabilized self-dual heterotic string
models is reviewed in~\cite{rev}, which includes extensive references 
to, and comparisons with, other models for dilaton stabilization.

Reviews of other models for communicating SUSY breaking from the
hidden sector to the observable sector can be found in summer school
and conference proceedings such as SUSY 07.  All of these models
predict different patterns of soft SUSY breaking.  Typically these
patterns appear simple at some high scale, such as the Planck scale,
the scale of gauge coupling unification or the scale of supersymmetry
breaking itself; for example in the case of gaugino condensation
described above, supersymmetry is broken at the condensation scale
$\Lambda_c\sim10^{14}\GeV$.  As discussed by Peskin~\cite{peskin}, it is
these high energy parameters that inform us about physics at near
Planck-scale energies, that we want to determine, while it is the low
energy parameters that we will measure.  Just as we were able to use
the renormalization group equations of quantum field theory to
determine high energy values of the couplings and discover near
unification in the SM and more accurate unification in the MSSM, we
can use similar QFT tools to probe high energy values of soft
parameters.  For example, gaugino masses scale with energy the same
way as the fine structure constants $\alpha_i(\mu) =
g^2_i(\mu)/4\pi^2$, so if they have a common mass $m$ at the
unification scale, at the scales $\mu$ at which the masses are
measured they will be in the ratio $m_1(\mu):m_2(\mu):m_3(\mu) =
\alpha_1(\mu):\alpha_2(\mu):\alpha_3(\mu)$.  Relations among squark
and slepton masses are more complicated.  Since squarks have strong QCD
couplings which increase their masses as $\mu$ decreases, they are
expected to be heavier than sleptons if all the sfermions have the
same mass at some high scale.  The most massive third family has
larger couplings to the Higgs field, which has the opposite effect,
making them lighter than their companions with the same gauge charges.
The gaugino and Higgsino sector is even more complicated with very
model-dependent mixing among the neutralinos discussed in \mysec{dark},
and among the ``charginos'' $\tilde W^{\pm}, H^{\pm}_u,H^{\pm}_d$.
For more details on the predicted spectra for different models the
reader is referred to the reviews~\cite{peskin} and~\cite{rev}.

\mysection{Concluding remarks}\label{conc}
We are very optimistic that the MSSM, or some extension of it, for
which there are many proposals in the literature (the simplest being
the nMSSM which postulates an additional scalar
superfield~\cite{nmssm}), will correctly describe the particle and
superparticle spectrum that will be produced at LHC energies.  We know
that our friend Julius shared this conviction.  Clearly the MSSM and
its extensions still have new difficulties of their own, even though
they solve problems of the SM as we have attempted to explain. The
history of physics teaches us that this is a situation which often
occurs when an important step forward takes place in our understanding
of the basic equations governing the fundamental processes of our
universe.

If the superpartners are not found at the LHC, it could simply be that
the accelerator energy is just not high enough and that their masses
are beyond the reach of the LHC. Even if the superpartners are found at
the LHC, it will be very difficult to distinguish among different
models, and even among SUSY and other proposals such as ``large extra
space-time dimensions''~\cite{5d}.  One will need a higher {\it
  precision} instrument such as the International Linear Collider
(ILC) to resolve these issues.  Will there be the political will to
build it? Unfortunately, we are not optimistic about this.

Another possibility is that the beautiful mathematics of supersymmetry
will find its application to physics in a different interpretation than
the one assumed so far.  To explain what we mean, we refer to the
history of non-Abelian gauge theories, which for a long time were
thought to be only a description of vector resonances.  It took the
realization of their property of renormalizability {\it as well as} a
total paradigm change, due to the development of the quark picture and
of the flavor-color properties of gauge theories, which led to the
very successful SM.  The analogy is striking for us.  As we explained,
SUSY quantum gauge field theories are renormalizable and have even
fewer divergences than generic gauge field theories.

Finally SUSY, through supergravity, provides a unified understanding
of all forces of nature, although not a perfect one, neither at the
four dimensional quantum field theory level (see however~\cite{zvi}),
nor at the superstring level.

\vskip 0.20in
\noindent {\bf \Large Acknowledgments}

\vspace{5pt}

\noindent 
This work was supported in part by the Director, Office of Science,
Office of High Energy and Nuclear Physics, Division of High Energy
Physics of the U.S. Department of Energy under Contract
DE-AC02-05CH11231, in part by the National Science Foundation under
grant PHY-0457315.

\end{document}